# Watt-level single-frequency tunable Neodymium MOPA fiber laser operating at 915-937 nm

S. Rota-Rodrigo,[1,*] B. Gouhier,[1] M. Laroche,[2] J. Zhao,[1] B. Canuel,[1] A. Bertoldi,[1] P. Bouyer,[1] N. Traynor,[3] B. Cadier,[4] T. Robin,[4] And G. Santarelli[1]

[1]LP2N, Univ. Bordeaux – CNRS – Institut d'Optique Graduate School, F-33400 Talence, France
[2]CIMAP, ENSICAEN, CNRS, CEA/IRAMIS, Université de Caen, 14050 Caen cedex, France
[3]Azur Light Systems, Pessac, France
[4]IXBLUE Photonics, Rue Paul Sabatier, Lannion, France
*Corresponding author: Sergio.rota@institutoptique.fr

We have developed a Watt-level single-frequency tunable fiber laser in the 915-937 nm spectral window. The laser is based on a neodymium-doped fiber master oscillator power amplifier architecture, with two amplification stages using a 20 mW extended cavity diode laser as seed. The system output power is higher than 2 W from 921 to 933 nm, with a stability better than 1.4% and a low relative intensity noise.

Watt-level Single-Frequency (SF) lasers have been thoroughly investigated due to their widespread applications ranging from atom cooling [1], to coherent LIDAR [2], and laser spectroscopy [3], among others. However, to date there are still uncovered wavelength domains due to technological challenges, especially in the 900-940 nm window, as well as the corresponding frequency-doubled 450-470 nm spectral domain, for applications such as laser-cooling of atoms [4], high resolution 3D lithography [5], and underwater communications [6]. Recently, neodymium (Nd) doped fiber operating on the $^4F_{3/2}$-$^4I_{9/2}$ transition have attracted high interest for the development of high-power lasers operating in the 900-940 nm spectral window. However, this type of lasers has only been demonstrated in longitudinal multimode operation, by using resonant cavity configurations combined with wavelength selection elements [7-9] or with photonic crystal fiber [10], in order to increase the gain at 910-940 nm and suppress the 1060 nm emission. In this Letter, we report for the first time to the best of our knowledge a Watt-level single-frequency all-fiber laser, tunable in the 915-937 nm window, with an output power in excess of 2 W from 921 to 933 nm.

The Master Oscillator Power Amplifier (MOPA) Nd-doped fiber laser is composed of two amplification stages, a preamplifier and a booster amplifier (see Fig. 1), seeded by a low power (20mW) SF narrow linewidth (<100 kHz) Extended Cavity Diode Laser (ECDL, Toptica DL Pro 940). This double stage configuration is compatible with low- power sources (<5 mW), like fiber-based sources based on distributed Bragg reflector (DBR) lasers [11].

The gain medium is based on a special designed Nd-doped double-clad fiber with core/cladding diameters of 5 μm/125 μm, NA=0.12 and clad absorption of 0.8 dB/m at λ=808 nm (iXblue Photonics). Standard Nd-doped fibers usually present a very high gain on the $^4F_{3/2}$-$^4I_{11/2}$ transition (near 1060 nm) due to its true four-level nature, which prevents laser operation near 900 nm. The fiber we adopted is designed with a W-type core refractive index profile in order to suppress the stimulated emission at 1060 nm by bending induced losses. Two Nd-fiber sections of 5m, coiled with a diameter of 6 cm, were used to implement the preamplifier and the booster amplifier. The preamplifier was pumped with a 4W laser diode at 808 nm via a multimode (MM) combiner. The booster amplifier was instead pumped with a 25 W pump laser diode at 808 nm through a high-power (HP) MM combiner.

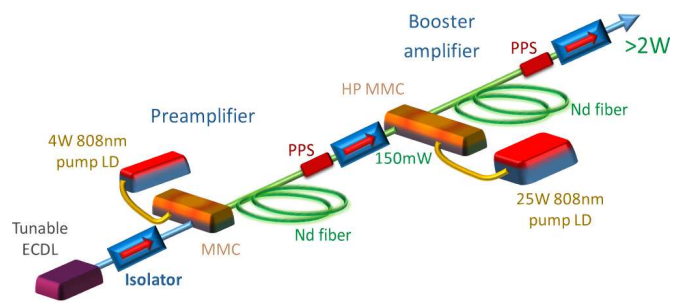

Fig. 1. Schematic of the tunable single-frequency MOPA laser. (MMC: Multi-Mode Combiner, PPS: in-fiber Pump Power Stripper)

To avoid stray reflections and achieve stable operation of the amplification stages, three polarization-insensitive fiber-coupled isolators were used as shown in Fig. 1. Also, the output fiber end-face was angle cleaved. Due to the low absorption of the Nd-doped fiber, the residual pump after each stage was removed by in-fiber pump power strippers.

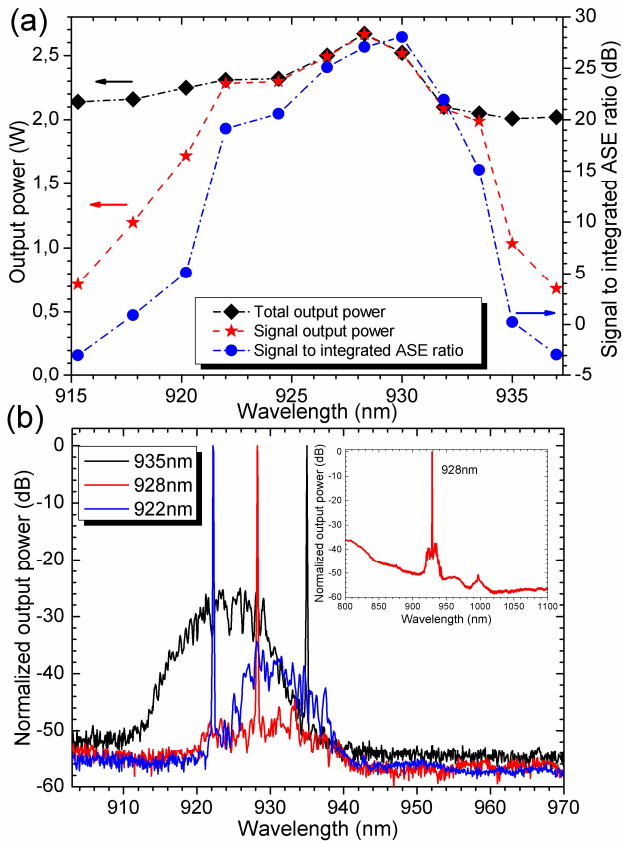

*Fig. 2. (a) Total output power (black), signal output power (red), and signal to integrated ASE ratio (blue) versus seed wavelength. (b) Optical spectra of the MOPA laser for different wavelengths measured with an OSA (resolution of 0.1 nm). (inset) Optical spectrum of the MOPA laser at 928 nm, from 800 to 1100 nm (resolution 0.5 nm)*

The tunable operation of the MOPA laser is shown in Fig. 2(a). The black trace represents the total output power measured with a power-meter. This power is the combination of signal and Amplified Spontaneous Emission (ASE), as can be seen in the inset of Fig. 2 (b). This spectrum illustrates the effective suppression of 1060 nm signal, due to the fiber design, as well as the absence of any residual pump at 808 nm, thanks to the in-fiber pump power strippers. The blue curve shows the ratio between the signal and the integrated ASE powers. When the seed wavelength is tuned away from the maximum emission window of the Nd-doped fiber, the signal amplification is reduced and the ASE of the system is thus increased. This signal to integrated ASE degradation can be seen Fig. 2(b), which shows the output optical spectrum of the MOPA fiber laser operating at different input seed wavelengths. We reach signal to integrated ASE ratios over 20 dB from 923 nm to 932 nm, with a maximum of 28 dB at 930 nm.

The actual power of the signal corresponds to the red trace of Fig. 2(a). The laser presents a tunable range over 22 nm, with an output power in excess of 1 W in the range of 917-935 nm, and more than 2 W from 921 nm to 933 nm. A maximum output power of 2.6 W was achieved at 928 nm.

Figure 3 shows the evolution of the total output power of the MOPA laser system versus the booster amplifier pump power. The slope efficiency with respect to the incident pump power was about 11% from 922 nm to 931 nm, and was lower out of this band due to the presence of ASE. Considering that only 60% of the pump power is absorbed, the optical-to-optical efficiency was higher than 18%, which is in agreement with the results reported in [12]. The preamplifier optical-to-optical efficiency, operated at a constant pump power of 4W, was about 7%. The low efficiency can be explained by the neodymium ions clustering, due to a high concentration of $Nd^{3+}$: such effect causes a strong decrease in the efficiency of the laser conversion [12].

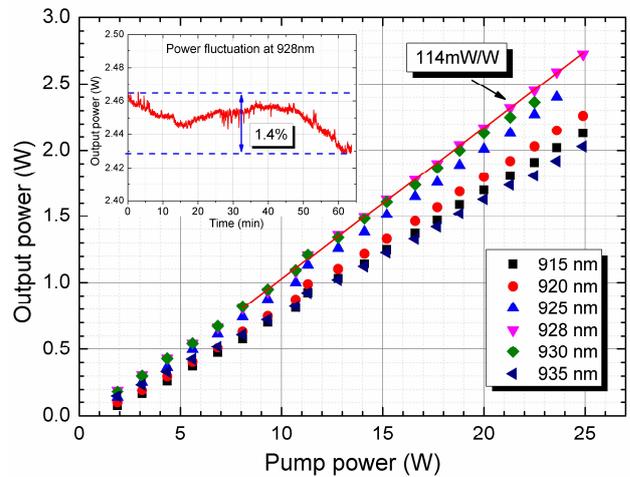

*Fig. 3. Laser total output power versus booster amplifier pump power for different seed wavelengths. (Inset) Power stability at 928 nm on a 1 hour interval.*

The output power level was monitored for 1 hour in order to study the output optical power stability of the laser, for a nominal power of ~2.5 W at 928 nm. The result (Fig. 3 Inset) shows a peak-to-peak power fluctuation lower than 1.4%. The long term fluctuations could be explained in terms of temperature changes of the room.

In Fig. 4(a), we show the measured Relative Intensity Noise (RIN) for the different stages of the laser. The ECDL seed exhibits a very low intensity noise approaching -160 dBc/Hz beyond a few tens of kHz. The RIN starts from -120 dBc/Hz at low frequency and after rolling-off at 20 dB/dec it reaches -158 dBc/Hz at a few hundreds of kHz. In the same way, the RIN of the MOPA laser is dominated by the booster amplifier pump noise. In this case, the RIN shows a slope of 40 dB/dec, which is well explained by the combination of the 20 dB/dec of the pump noise behavior with the first-order low pass filter from the pump transfer function. In this case, the RIN of the laser at 2.5 W output power is -122 dBc/Hz at low frequency and decreases to -158 dBc/Hz at high frequency.

One of the major limiting effects in high power lasers is Stimulated Brillouin Scattering (SBS), a non-linear process that adds noise in the system at high frequency. Monitoring the laser

RIN up to 10 MHz can be used as an efficient method for early detection of SBS [13,14]. From Fig. 4, we can see that excess noise at high frequency (>1 MHz) is absent, when the laser operates at 2.5 W output power. Therefore, the MOPA laser system is operating below SBS threshold. However, we observed that for a longer fiber, high frequency noise increases because of the close SBS threshold.

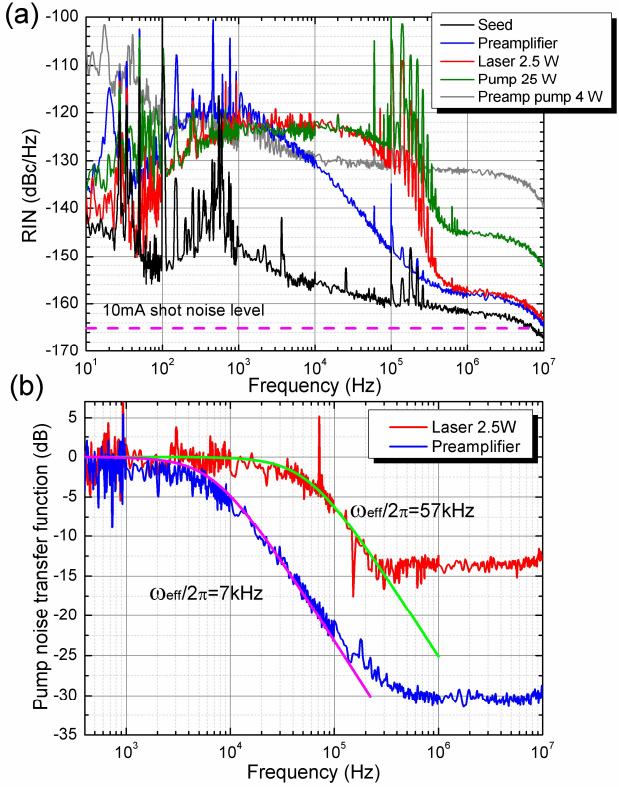

*Fig. 4. (a) RIN on the 10 Hz-10 MHz frequency window for: seed laser at 928 nm and 20 mW output power (black line); preamplifier with 150 mW output power (blue line); full MOPA laser with 2.5 W output power (red line); pump laser with 25 W (green line). All measurements are taken with a photodetector current of about 10 mA. (b) pump-to-signal noise transfer functions.*

The dynamics of the noise added by the amplification process has been investigated by several groups for Erbium and Ytterbium amplifiers [15-17]. It has been shown that in the case of low noise seed sources, the main contributor to the output noise is the pump noise. The latter couples to the output signal via a transfer function which can be modeled by a first order low pass filter with the corner frequency determined by fiber parameters and output power. Our aim is now to apply this approach to Neodymium doped fiber amplifiers by comparing the experimental frequency corner of our system, with the theoretical one derived from the model by using the fiber parameters reported in [18].

Figure 4(b) show the pump-to-signal transfer function for the preamplifier and the MOPA laser. The transfer function can be obtained by modulating the pump intensity as in [16, 17]. However, a simple way to retrieve it, consist in divide the measured output laser signal RIN by the pump RIN. To this end, we accurately measured the RIN of the multimode pump laser diodes used in both stages (Fig. 4(a)). We applied a first-order low-pass filter fit to the transfer functions, and we obtained the corner frequencies for the preamplifier (7±0.5 kHz) and MOPA laser (57±5 kHz).

From the gain dynamics model [16] the corner frequency can be expressed as:

$$\omega_{eff} = P_s^0(L) \cdot B_s + P_p^0(L) \cdot B_p + 1/\tau \quad (1)$$

where $P_s^0(L)$ and $P_p^0(L)$ are the seed and pump powers at the output of the gain fiber (z=L) in the units of number of photon per second, $\tau$ the fluorescence lifetime in the $^4F_{3/2}$-$^4I_{9/2}$ transition of Nd in seconds and, $B_s$ and $B_p$ the coupling factors of the signal and the pump, respectively. The coupling factors are defined as

$$B_s = \Gamma_s(\sigma_s^a + \sigma_s^e)/A \quad (2)$$
$$B_p = \Gamma_p \sigma_p^a/A \quad (3)$$

where $\Gamma_s$ and $\Gamma_p$ are the signal and pump overlap with the doped region, $\sigma_s^a$ and $\sigma_s^e$ the absorption and emission signal cross sections, $\sigma_p^a$ the pump absorption cross sections, and $A$ the fiber mode field area.

In MOPA systems with low residual pump [17], the corner frequency can be approximated to $P_s^0(L) \cdot B_s$. However, our Nd-MOPA system presents a signal to residual pump ratio of 1:4 that must be taken into account in equation (1). By using this model and the values shown in Table 1, we derive the theoretical corner frequency of our Nd fiber.

**Table 1. Gain dynamics model parameters**

| Parameter (units) | Value |
|---|---|
| Core diameter (μm) | 5±0.5 |
| Signal cross section absorption $\sigma_s^a$ (m²) from [18] | 2x10⁻²⁶ |
| Pump cross section absorption $\sigma_p^a$ (m²) from [18] | 1.5x10⁻²⁴ |
| Signal cross section emission $\sigma_s^e$ (m²) from [18] | 3x10⁻²⁵ |
| Signal overlap with the doped region $\Gamma_s$ | 0.8 |
| Pump overlap with the doped region $\Gamma_p$ | 0.0021 |
| Fluorescence lifetime in the $^4F_{3/2}$-$^4I_{9/2}$ transition of Nd (s) | 4.75x10⁻⁴ |
| $P_s^0(5m)$ (W) | 2.5 |
| $P_p^0(5m)$ (W) | 10 |
| Theoretical corner frequency (kHz) | 160±30 |
| Experimental corner frequency (kHz) | 57±5 |

The calculated corner frequency value (160±30 kHz) differs from the measured one by nearly a factor 3. This discrepancy is likely to be due to the uncertainty on the spectroscopy data reported in [18]. Precise spectroscopy measurements of Nd-doped fibers in the $^4F_{3/2}$-$^4I_{9/2}$ level transition are in progress.

In conclusion, a tunable single-frequency neodymium MOPA fiber laser with more than 2 W output power in the 921-933 nm window has been demonstrated. The laser is tunable from 915 nm to 937 nm, and achieves signal-to- integrated ASE ratios up to 28 dB. The laser also exhibits a stability better than 1.4% in one hour, as well as a very low RIN. In the future, we plan to develop an all-fiber polarized MOPA laser which will enable the generation of blue laser radiation, by cavity-assisted second harmonic generation process. We also plan to increase the output power up to 10 W by using Large Mode Area fibers (LMA) [7,10]. However, the development of a mode field adapter with Nd doped fibers will be required to maintain an all fiber structure.

**Acknowledgment.** We acknowledge the financial support of the cluster of excellence Lasers and Photonics in Aquitane (LAPHIA) in the framework of Programme d'Investissements d'Avenir (PIA) (project MIGA-PHYS), l'Agence Nationale de la Recherche (ANR) program LabCom 2014 and ALCALINF, le Conseil Régional d'Aquitaine (project IASIG-3D). This project has received funding from the European Union's Horizon 2020 research and innovation programme under the Marie Sklodowska-Curie grant agreement No 748839.

**Link:** https://doi.org/10.1364/OL.42.004557
**Citation:**
S. Rota-Rodrigo, B. Gouhier, M. Laroche, J. Zhao, B. Canuel, A. Bertoldi, P. Bouyer, N. Traynor, B. Cadier, T. Robin, and G. Santarelli, "Watt-level single-frequency tunable neodymium MOPA fiber laser operating at 915–937 nm," Opt. Lett. **42**, 4557-4560 (2017)